\documentclass[12pt]{article}
\usepackage{graphicx}
\usepackage{amssymb,epsfig}
\usepackage{lscape,graphics,amsmath}
\usepackage{latexsym,amsfonts}
\usepackage{axodraw}
\usepackage[english]{babel}

\newcounter{diags1}
\newcounter{diags2}
\newcounter{diags3}
\begin{document}

\begin{center}
{\Large \bf Neutrino oscillation processes with a change of lepton  flavor in quantum field-theoretical approach}\\
\vspace{4mm} Vadim O. Egorov$^{1,2}$, Igor P.~Volobuev$^1$\\
\vspace{4mm} $^1$Skobeltsyn Institute of Nuclear Physics, Moscow State
University
\\ 119991 Moscow, Russia\\
$^2$Faculty of Physics, Moscow State University, 119991 Moscow, Russia
\end{center}

\vspace{0.5cm}
\begin{abstract}
The oscillating probabilities of lepton flavor changing neutrino
oscillation processes, where neutrinos are detected by
charged-current and neutral-current interactions, are calculated
in a quantum field-theoretical approach to neutrino oscillations
based on a modification of the Feynman propagator in the momentum
reprsentation. The approach is most similar to the standard
Feynman diagram technique in the momentum representation. It is
found that the oscillating distance-dependent probabilities of
detecting an electron in experiments with neutrino production in
the  muonic decay of $\pi^+$-meson  and the detection of the
produced neutrino by charged-current and neutral-current
interactions exactly coincide with the corresponding probabilities
calculated in the standard approach.
\end{abstract}

\section{Introduction}
Neutrino oscillations are an experimentally confirmed phenomenon
that is widely discussed in theoretical physics. It is usually
interpreted as the transition from a neutrino flavor state to
another neutrino flavor state depending on the distance traveled
\cite{Giunti:2007ry,Bilenky:2010zza,Petcov}. This
interpretation is based on the standard quantum mechanical
description of neutrino oscillations, where the neutrino flavor
states are assumed to be superpositions of states with definite
masses described by plane waves, and it is postulated that it is
these flavor states that are produced in weak interactions.
However, in local quantum field theory 4-momentum is conserved in
any interaction vertex, which leads to different neutrino
mass-eigenstate components of a flavor state having different
momenta and energies. As a result, there is a problem with
violation of  energy-momentum conservation, which was extensively
discussed in the literature (see, e.g.
\cite{Giunti:1993se,Grimus:1996av,Beuthe:2001rc,Cohen:2008qb,Lobanov:2015esa}).

A solution to the problem can be found by considering off-shell
neutrinos.  The idea to treat the neutrino mass eigenstates as
virtual particles and to describe their motion to a detection
point by the Feynman propagators was first put forward in paper
\cite{Giunti:1993se}. Later this approach was developed in papers
\cite{Grimus:1996av,Beuthe:2001rc}. In this approach neutrino
oscillations occur as a result of interference of the amplitudes
of  processes due to all the three intermediate  virtual neutrino
mass eigenstates.  However, the calculations  of the amplitudes in
this approach are essentially different from the standard
calculations in the  Feynman diagram technique in the momentum
representation. This is due to the standard S-matrix formalism of
QFT, which is not convenient for describing processes at finite
distances and finite time intervals.  To describe a localization
of particles or nuclei, which produce and detect neutrinos,  one
has to use wave packets, which makes the calculations rather
complicated.

In paper \cite{Volobuev:2017izt} a modified perturbative S-matrix
formalism was put forward, which allows one to consistently
describe the neutrino oscillation processes in the framework of
quantum field theory using only plane waves. The formalism
is based on the Feynman diagram technique in the coordinate
representation \cite{Feynman:1949zx} supplemented by modified
rules of passing to the momentum representation. The calculation
procedure proper is very similar to the Feynman diagram technique
in the momentum representation, where we make use of a modified
Feynman propagator. The approach was  developed in paper
\cite{Egorov:2017qgk}, where we have shown explicitly that the
suggested formalism  exactly reproduces the results of the
standard approach in the case, where neutrinos (together with
positrons) are produced in the charged-current interaction with
nuclei and detected in both neutral-current and charged-current
interactions with electrons.

In the present paper we will use the modified S-matrix formalism
to calculate probabilities of neutrino oscillation processes
non-diagonal in the lepton flavors. Namely, we will consider the
processes, where a neutrino is produced in the muonic decay of a
$\pi^+$-meson and detected in the neutral-current and
charged-current interactions with  electrons. We will show that
the results of our approach again exactly coincide with what one
expects in the standard approach.

\section{Oscillations in experiments with neutrino detection
by neutral-current and charged-current interactions}

The minimal extension of the Standard Model (SM) by the right
neutrino singlets is considered. After the diagonalization of the
terms sesquilinear in the neutrino fields, the charged-current
interaction Lagrangian of leptons takes the form
\begin{equation}\label{L_cc}
L_{cc} = - \frac{g }{2\sqrt{2}}\left(\sum_{i,k = 1}^3 \bar l_i
\gamma^\mu (1 - \gamma^5)U_{ik}\nu_k   W^{-}_\mu + h.c.\right),
\end{equation}
where $l_i$ denotes the field of the charged lepton of the i-th
generation,  $\nu_i$ denotes the field of the neutrino mass
eigenstate most strongly coupled to $l_i$ and $U_{ik}$ stands for
the Pontecorvo-Maki-Nakagawa-Sakata (PMNS) matrix.

We are going to consider the process, where a neutrino is produced
in the decay of  $\pi^+$-meson and is detected by the
charged-current and neutral-current interactions with  electrons.
Due to the structure of the interaction Lagrangian, the process is
represented in the lowest order by the following two diagrams:
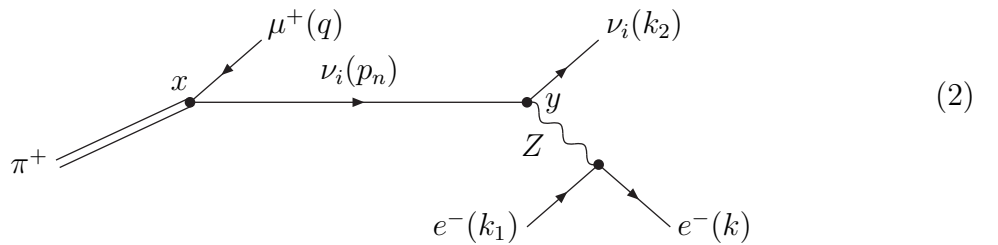
\vspace{1cm}
\begin{center}
\begin{picture}(193,81)(0,0)
\Text(70.0,94.0)[l]{$\mu^+ ( q)$}\ArrowLine(67.5,88.0)(40.5,64.5)
\Text(40,72)[r]{$x$} \Line(-10,42.7)(40.5,66.2) \Line(-8.6,39.9)(42,63.5)
\Text(-13,42.5)[r]{$\pi^+$}
\Vertex(40.5,64.5){2} \ArrowLine(40.5,64.5)(167.5,64.5) \Vertex
(167.5,64.5){2} \Text(104.8,70.5)[b]{$\nu_i ( p_n )$}
\ArrowLine(167.5,64.5)(194.5,88.0) \Text(198.0,94.0)[l]{$\nu_i ( k_2 )$}
\Text(175.0,64.5)[l]{$y$} \Text(166.0,48.5)[l]{$Z$}
\Photon(167.5,64.5)(194.5,41.0){2}{3.0} \Vertex (194.5,41.0){2}
\ArrowLine(167.5,17.5)(194.5,41.0) \Text(132.5,17.5)[l]{$e^- ( k_1 )$}
\ArrowLine(194.5,41.0)(221.5,17.5) \Text(225.0,17.5)[l]{$e^- ( k )$}
\Text(330.0,60.5)[b]{\addtocounter{equation}{1}(\arabic{equation})}
\setcounter{diags2}{\value{equation}}
\end{picture}
\end{center}

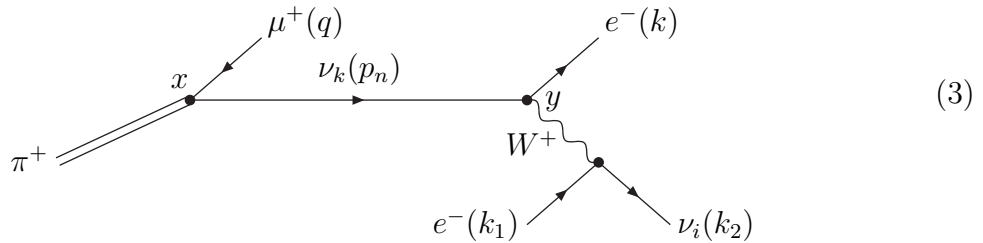
\vspace{1cm}
\begin{center}
\begin{picture}(193,81)(0,0)
\Text(70.0,94.0)[l]{$\mu^+ (q)$}\ArrowLine(67.5,88.0)(40.5,64.5)
\Text(40,72)[r]{$x$} \Line(-10,42.7)(40.5,66.2) \Line(-8.6,39.9)(42,63.5)
\Text(-13,42.5)[r]{$\pi^+$}
\Vertex(40.5,64.5){2} \ArrowLine(40.5,64.5)(167.5,64.5) \Vertex
(167.5,64.5){2} \Text(104.8,70.5)[b]{$\nu_k ( p_n )$}
\ArrowLine(167.5,64.5)(194.5,88.0) \Text(197.5,94.0)[l]{$e^- ( k )$}
\Text(175.0,64.5)[l]{$y$} \Text(160.0,48.5)[l]{$W^+$}
\Photon(167.5,64.5)(194.5,41.0){2}{3.0} \Vertex (194.5,41.0){2}
\ArrowLine(167.5,17.5)(194.5,41.0) \Text(132.5,17.5)[l]{$e^- ( k_1 )$}
\ArrowLine(194.5,41.0)(221.5,17.5) \Text(225.0,17.5)[l]{$\nu_i ( k_2 )$}
\Text(330.0,60.5)[b]{\addtocounter{equation}{1}(\arabic{equation})}
\setcounter{diags3}{\value{equation}}
\end{picture}
\end{center}
\vspace{-0.5cm}
In diagram (\arabic{diags3}) all the three virtual neutrino mass
eigenstates  contribute, so the corresponding amplitude should be
summed up over the index $k=1,2,3$. At the same time, both
diagrams have  neutrino mass eigenstate $\nu_i$  in the final
state, thus we should sum the resulting probability over $i$ to
get the probability of registering an electron.

Let us denote the 4-momenta of the particles as it is depicted in
the diagram:  the momentum of the antimuon is $q$, the momentum of
the virtual neutrinos is $p_n$, the momentum of the outgoing
electron is $k$, the momentum of the incoming electron  is $k_1$
and the momentum of the outgoing neutrino is $k_2$.

One can write out the amplitude in the coordinate
representation corresponding to diagrams
(\arabic{diags2})--(\arabic{diags3}) using the standard Feynman
rules formulated in textbook \cite{BOSH}. Next, according to the
prescription of the S-matrix formalism, in order to pass to the
momentum representation one would have to integrate the amplitude
with respect to $x$ and $y$ over  Minkowski space,  which means
that one considers the process to take place throughout  Minkowski
space-time and that the resulting probability of the process will
be the probability per unit volume and per unit time.

However, such an integration  would result in losing the
information about the space-time interval between the production
event and the detection event, because the experimental situation
in neutrino oscillation  experiments implies that the distance
between the production point and the detection point along the
neutrino propagation direction remains fixed. To generalize the
standard S-matrix formalism to the case of processes passing at
fixed distances, we  introduce the delta function $\delta(\vec
p(\vec y -\vec x)/|\vec p| - L)$ into the integral, $\vec p$
denoting the momentum of the virtual neutrinos and $L$ denoting
the distance between a neutrino source and a detector.  In so
doing we fix the distance between the production and detection
events, and only then we integrate the amplitude with respect to
$x$ and $y$ over  Minkowski space. Thus, just like in the standard
S-matrix formalism, we consider the process taking place
throughout Minkowski space-time, but the distance between the
production and detection events  along the momentum of the
neutrino beam is now fixed by the delta function. This is
equivalent to  replacing the standard Feynman fermion propagator
in the coordinate representation $S^c_i(y-x)$ by
$S^c_i(y-x)\delta(\vec p(\vec y -\vec x)/|\vec p| - L)$.

The Fourier transform of this expression gives us the so-called
distance-dependent propagator of the neutrino mass eigenstate
$\nu_i$ in the momentum representation
\cite{Volobuev:2017izt,Egorov:2017qgk}. However, in paper
\cite{Egorov:2017qgk} it was argued that this distance-dependent
propagator is inconvenient for calculations, because its inverse
Fourier transformation cannot be defined, if the momentum of the
Fourier transform coincides with the momentum of the virtual
neutrinos in the argument of the delta function, which is needed
to describe neutrino oscillation processes. To circumvent this
problem, we introduce a different delta function $\delta(y^0 - x^0
- T)$ into the integral, which fixes the time interval between the
production and detection events. Later we will be able to express
the time interval $T$ in terms of the distance traveled by the
neutrinos in accordance with the formula $T = L p^0/|\vec p|$,
which is often used in describing neutrino oscillation processes.

Now the introduction of the delta function is equivalent to the
replacement of the standard Feynman fermion propagator in the
coordinate representation $S^c_i(y-x)$ by $S^c_i(y-x)\delta(y^0 -
x^0 - T)$. In this case the Fourier transform gives us the
so-called time-dependent propagator of the neutrino mass
eigenstate $\nu_i$ in the momentum representation, defined by the
relation:
\begin{equation}\label{prop_T_mom}
S^c_i(p_n,T) = \int dx\, e^{ip_n x} S^c_i(x)\,  \delta(x^0 - T).
\end{equation}

This integral can be evaluated exactly
\cite{Volobuev:2017izt,Egorov:2017qgk}:
\begin{equation}
\label{spin_prop}S^c_i \left( {p_n,T} \right) = i \, \frac{{\hat p_n -
\gamma _0 \left( {p_n^0  - \sqrt {\left( {p_n^0 } \right)^2  +
m_i^2 - p_n^2} } \right) + m_i}}{{2\sqrt {\left( {p_n^0 }
\right)^2 + m_i^2 - p_n^2} }} \, e^{i\left( {p_n^0  - \sqrt {\left(
{p_n^0 } \right)^2 + m_i^2 - p_n^2} } \right)T}\,,
\end{equation}
where the standard notation $\hat p_n = \gamma_\mu p_n^\mu $ is
used. The inverse Fourier transformation of this time-dependent
propagator is well defined, which allows us to retain the standard
Feynman diagram technique in the momentum representation just by
replacing the Feynman propagator by the time-dependent propagator.

 In paper \cite{Grimus:1996av} it was shown  that virtual particles
propagating at large macroscopic distances (or, equivalently,
propagating over macroscopic times) are almost on the mass shell,
which means that $|p_n^2 - m_i^2|/ \vec p_n^{\,2} \ll 1$. Applying
also the approximation of small neutrino masses, i.e. keeping
the neutrino masses only in the exponential, we can explicitly
represent the time-dependent neutrino propagator in the momentum
representation in the form \cite{Egorov:2017qgk}
\begin{equation}\label{prop_T_mom_c}
S^c_i(p_n,T) = i\, \frac{\hat p_n } { 2p_n^0}\,e^{-i\frac{m_i^2 - p_n^2}{ 2 p_n^0} T}\,.
\end{equation}
This time-dependent propagator will be used in the calculations
below  replacing the standard Feynman propagator. Such a technical
simplicity is an evident advantage of the discussed approach.

Now we are in a position to write out the amplitudes corresponding
to diagrams (\arabic{diags2})--(\arabic{diags3}) in the case,
where the time difference $y^0 - x^0$ is fixed and equal to $T$.
We assume that the momentum transfer in the production and
detection processes is small, so that we can use the approximation
of Fermi's interaction. The pion decay vertex is described in
accordance with the formulas in \S 5 of textbook
\cite{Okun:1982ap}. The amplitude corresponding to diagram
(\arabic{diags2}) in the momentum representation looks like
\begin{eqnarray}
 M_{nc}^{(i)}  &=&  - \frac{{G_F^{\, 2} }}{{2p_n^0 }}\cos \theta_c\, f_\pi
  \varphi _\pi  m_{(\mu )} U_{2i}^ *  e^{ - i\frac{{m_i^2  - p_n^2 }}{{2p_n^0 }}T} \bar \nu _i
   \left( {k_2 } \right)\gamma ^\mu  \hat p_n \left( {1 + \gamma ^5 } \right)\upsilon \left( q \right)
    \times \label{1} \\
 & & \times \left[ {\left( { - \frac{1}{2} + \sin ^2 \theta _W } \right)\bar u\left( k \right)\gamma _\mu
 \left( {1 - \gamma ^5 } \right)u\left( {k_1 } \right) + \sin ^2 \theta _W \bar u\left( k \right)\gamma _\mu
 \left( {1 + \gamma ^5 } \right)u\left( {k_1 } \right)} \right] , \nonumber
\end{eqnarray}
where $\theta_c$ is the Cabibbo angle,  $f_\pi$ is the pion decay
constant of the dimension of mass, $\varphi _\pi$ is the
(constant) pion wave function, $m_{(\mu )}$ is the muon mass, and
we have already applied the 4-momentum conservation condition in
the production vertex. Here and below we drop the fermion
polarization indices.

Similarly, the amplitude corresponding to diagram
(\arabic{diags3}) summed over the type $k$ of the intermediate
virtual neutrino can be written out to be
\begin{eqnarray}
M_{cc}^{(i)}  &=& \frac{{G_F^{\, 2} }}{{2p_n^0 }}\cos \theta_c\,
f_\pi  \varphi _\pi  m_{(\mu )} U_{1i}^ *
 \sum\limits_{k = 1}^3 {U_{1k} U_{2k}^ *  e^{ - i\frac{{m_k^2  - p_n^2 }}{{2p_n^0 }}T} } \times \nonumber \\
& & \times \bar \nu _i \left( {k_2 } \right)\gamma _\mu  \left( {1 - \gamma ^5 } \right)
u\left( {k_1 } \right)\bar u\left( k \right)\gamma ^\mu  \hat p_n \left( {1 + \gamma ^5 } \right)
\upsilon \left( q \right). \label{2}
\end{eqnarray}
It is convenient to use the Fierz identity to transpose  the
spinors $\bar u \left( k \right)$ and $\bar \nu_i \left( k_2
\right)$ in the latter amplitude, which makes it look similar to
the former one. We also introduce the following notations for the
time-dependent factors:
\begin{equation}\label{3}
A_i  = U_{2i}^ *  e^{ - i\frac{{m_i^2  - p_n^2 }}{{2p_n^0 }}T} ,
\qquad B_i  = U_{1i}^ *  \sum\limits_{k = 1}^3 {U_{1k} U_{2k}^ *
 e^{ - i\frac{{m_k^2  - p_n^2 }}{{2p_n^0 }}T} } .
\end{equation}
The total amplitude of the process with the neutrino $\nu_i$ in
the final state, which is the sum of the amplitudes $M_{nc}^{(i)}$
and $M_{cc}^{(i)}$, takes the form
\begin{eqnarray}
\! \! \! \! \! \! \! \! \! \! \! \! & & M_{tot}^{(i)}  =  -
\frac{{G_F^{\, 2} }}{{2p_n^0 }}\cos \theta_c\, f_\pi  \varphi _\pi
m_{(\mu )}
  \bar \nu _i \left( {k_2 } \right)\gamma ^\mu  \hat p_n \left( {1 + \gamma ^5 } \right)
  \upsilon \left( q \right) \times  \label{amp_tot} \\
\! \! \! \! \! \! \! \! \! \! \! \! & & \times \left[ {\left( {B_i  + A_i \left( { - \frac{1}{2} + \sin ^2 \theta _W } \right)} \right)
 \bar u\left( k \right)\gamma _\mu  \left( {1 - \gamma ^5 } \right)u\left( {k_1 } \right) +
 A_i \sin ^2 \theta _W \bar u\left( k \right)\gamma _\mu  \left( {1 + \gamma ^5 } \right)
 u\left( {k_1 } \right)} \right] . \nonumber
\end{eqnarray}
One can notice that the dimension of this amplitude is not usual.
Formally, it corresponds to the process, in which the time
difference $y^0 - x^0$ between the production and the detection is
exactly equal to $T$. However, in reality, a registration process
is not instant, it takes some time $\Delta t$, ${{\Delta t}
\mathord{\left/ {\vphantom {{\Delta t} {T}}} \right.
\kern-\nulldelimiterspace} {T}} \ll 1$. To find the amplitude of
the process  with  the registration time $\Delta t$ we have to
integrate amplitude (\ref{amp_tot}) with respect to $T$
 from $T-{{\Delta t} \mathord{\left/
{\vphantom {{\Delta t} {2}}} \right. \kern-\nulldelimiterspace}
{2}}$ to $T+{{\Delta t} \mathord{\left/ {\vphantom {{\Delta t}
{2}}} \right. \kern-\nulldelimiterspace} {2}}$. After dropping the
terms of the order ${{\Delta t}/ {T}}$, which are negligibly
small, the integration  results in the multiplication of the
amplitude by $\Delta t$. Hence,  expression (\ref{amp_tot}) should
be understood as the amplitude per unit time.

Our next step is to calculate the squared amplitude, averaged
with respect to the polarizations of the incoming particles and
summed over the polarizations of the outgoing particles. The
operation of averaging and summation will be denoted by angle
brackets. Applying again the approximation of small masses of
almost real intermediate neutrinos, $p_n^2=0$, we find that the
squared amplitude factorizes as follows:
\begin{eqnarray}
 \left\langle {\left| {M_{tot}^{(i)} } \right|^2 } \right\rangle
 &=& \left\langle {\left| {M_1 } \right|^2 } \right\rangle
 \left\langle {\left| {M_2^{(i)} } \right|^2 } \right\rangle
 \frac{1}{{4\left( {p_n^0 } \right)^2 }}, \label{sqr_amp_tot} \\
 \left\langle {\left| {M_1 } \right|^2 } \right\rangle
 &=& 4G_F^{\, 2} \cos ^2 \theta_c\, f_\pi ^2 m_{(\mu )}^2 \left( {p_n q} \right), \label{sqr_amp_1} \\
 \left\langle {\left| {M_2^{(i)} } \right|^2 } \right\rangle
 &=& 64G_F^{\, 2} \left[ {\left| {B_i  + A_i \left( { - \frac{1}{2} +
 \sin ^2 \theta _W } \right)} \right|^2 \left( {k_1 p_n } \right)^2  +
 \left| {A_i } \right|^2 \sin ^4 \theta _W \left( {kp_n } \right)^2  - } \right. \nonumber \\
 & & \left. { - \left( {{\mathop{\rm Re}\nolimits} \left( {A_i B_i ^ *  } \right) +
 \left| {A_i } \right|^2 \left( { - \frac{1}{2} + \sin ^2 \theta _W } \right)} \right)
 \sin ^2 \theta _W m^2 \left( {k_2 p_n } \right)} \right], \label{sqr_amp_2}
\end{eqnarray}
where $\left\langle {\left| {M_1 } \right|^2 } \right\rangle$ is
the squared amplitude of the decay process of  $\pi^+$-meson into
antimuon and a massless fermion, $\left\langle {\left| {M_2^{(i)}
} \right|^2 } \right\rangle$ is the squared amplitude of the
scattering process of a massless fermion and the initial electron,
$m$ standing for the electron mass.

Let us denote the 4-momentum of the decaying pion by $p_\pi$  and
the 4-momentum of the neutrinos to be detected by $p$. The
experimental setting defines that the momentum $\vec p$ is
directed from a source to a detector and satisfies the momentum
conservation condition $\vec p_\pi - \vec q - \vec p = 0$ in the
production vertex. In other words, $\vec p$ is a special value of
$\vec p_n$, which is directed from the source to the detector.
Actually, the selection of the single value $\vec p$ of the
momenta of the neutrinos to be detected is an approximation, which
is applicable, when the distance between the source and the
detector is much larger than their  sizes. We also recall that we
work in the approximation $p^2=0$. Following the prescription
formulated in paper \cite{Egorov:2017qgk}, in order to find the
differential probability of the process one should multiply the
amplitude $\left\langle {\left| {M_{tot}^{(i)} } \right|^2 }
\right\rangle$ by the delta function of  energy-momentum
conservation $(2 \pi)^4 \delta ( p_\pi + k_1 - q - k - k_2 )$ and
by the delta function $2 \pi \delta ( p_\pi - q - p )$, which
selects the momentum of the intermediate neutrinos, substitute $p$
instead of $p_n$ and integrate the result with respect to the
momenta of the final particles, namely antimuon, electron and
neutrino, in accordance with the standard rules of probability
calculations. The factor $2\pi$ in front of the latter delta
function arises after an averaging over the momenta of the
neutrinos to be detected, which, because of non-zero sizes of the
source and the detector, really lie inside a small cone with the
axis along the vector $\vec p$.

Due to the factorization of the squared amplitude the differential
probability factorizes as follows:
\begin{eqnarray}
 \frac{{dW^{(i)} }}{{d\vec p}} &=& \frac{{dW_1 }}{{d\vec p}}\,W_2^{(i)} , \label{prob_tot} \\
 \frac{{dW_1 }}{{d\vec p}} &=& \frac{1}{{2p_\pi ^0 }}\frac{1}{{\left( {2\pi } \right)^3 2p^0 }}
 \int {\frac{{d^3 q}}{{\left( {2\pi } \right)^3 2q^0 }}
 \left\langle {\left| {M_1 } \right|^2 } \right\rangle
 \left( {2\pi } \right)^4 \delta \left( {p_\pi   - q - p} \right)} , \label{prob_1} \\
 W_2^{(i)}  &=& \frac{1}{{2p^0 2k_1^0 }}\int {\frac{{d^3 k}}{{\left( {2\pi } \right)^3 2k^0 }}\frac{{d^3 k_2 }}{{\left( {2\pi } \right)^3 2k_2^0 }}\left\langle {\left| {M_2^{(i)} } \right|^2 } \right\rangle } \left( {2\pi } \right)^4 \delta \left( {k_1  + p - k - k_2 } \right). \label{prob_2}
\end{eqnarray}
Here $\frac{{dW_1 }}{{d\vec p}}$ is the differential probability
of the $\pi$-meson decay  into an antimuon and a massless fermion
with the fixed momentum $\vec p$,\, $W_2^{(i)}$ is the probability
of the scattering  process of  electron and a massless fermion
with the production of an electron and  neutrino mass eigenstate
$\nu_i$.

In order to find the total differential probability of detecting
an electron in the final state we have to sum the differential
probability $\frac{{dW^{(i)} }}{{d\vec p}}$ over $i=1,2,3$. Since
$\frac{{dW_1 }}{{d\vec p}}$ does not depend on $i$,  we should sum
only $W_2^{(i)}$; the result will be denoted by $W_2$. Then the
total differential probability of detecting an electron in the
final state can be written as
\begin{equation}
 \frac{{dW} }{{d\vec p}}  = \frac{{dW_1 }}{{d\vec p}}\,W_2 . \label{prob_tot1}
\end{equation}

Since the momentum $p_n$ of the intermediate virtual neutrinos  is
now fixed and equal to $p$, we can substitute $T = {{Lp^0 }
\mathord{\left/
 {\vphantom {{Lp^0 } {\left| {\vec p} \right|}}} \right.
 \kern-\nulldelimiterspace} {\left| {\vec p} \right|}}$ into all
 the formulas from now on. This substitution is consistent,
 because the neutrinos are almost on the mass shell, and $|\vec p|/p^0$
 can be considered as the neutrino speed with a very high accuracy.

Next we observe that  the experimental setting fixes only the
direction of the neutrino momentum,  but not its length $\left|
{\vec p_n} \right| = \left| {\vec p} \right|$. Therefore, to find
the probability of the process we must also integrate
(\ref{prob_tot1}) with respect to $\left| {\vec p} \right|$ over
all the admissible values. The maximal value of $\left| {\vec p}
\right|$ is determined by the production process and the minimal
one is determined by the detection process. Here the production
process is a two-body decay, which means that the lengths of the
neutrino and antimuon momenta are already fixed by energy-momentum
conservation. It results in $\frac{{dW_1}}{{d\vec p}}$ being
singular, and this singularity  is eliminated  by the integration.
The final result for the probability of the process is as follows:
\begin{equation}\label{prob_tot_int}
\frac{{dW}}{{d \Omega}} = \int { \sum\limits_{i = 1}^3
{\frac{{dW^{(i)}}}{{d\vec p}}} \left| {\vec p} \right|^2 d\left|
{\vec p} \right| } = \frac{{dW_1}}{{d \Omega}} \left. {W_2 }
\right|_{\left| {\vec p} \right| = \left| {\vec p} \right|^
*  } ,
\end{equation}
where
\begin{equation}\label{dif_prob_prod}
\frac{{dW_1 }}{{d\Omega }} = \frac{{G_F^2 \cos ^2 \theta_c\, f_\pi
^2 }}{{8\left( {2\pi } \right)^2 }} \frac{{m_{(\mu )}^2 \left(
{m_\pi ^2  - m_{(\mu )}^2 } \right)^2 }}{{p_\pi ^0 \left( {p_\pi
^0  - \left| {\vec p_\pi  } \right|\cos \theta  } \right)^2 }}
\end{equation}
is the differential probability of the $\pi$-meson decay into  an
antimuon and a massless fermion with the fixed direction of the
momentum, and
\begin{equation}\label{p^star}
\left| {\vec p} \right|^ *   = \frac{{m_\pi ^2  - m_\mu ^2 }}
{{2\left( {p_\pi ^0  - \left| {\vec p_\pi  } \right|\cos \theta
 } \right)}};
\end{equation}
the coordinate system is chosen in such a way that the pion
momentum $\vec p_\pi$ is directed along the $Z$-axis, and $\theta$
is the polar angle of $\vec p$. After all these transformations
the probability (\ref{prob_tot_int}) can be interpreted as the
probability per unit length of the  source  and per unit length of
the detector.

As one can see, differential probability (\ref{dif_prob_prod}) has
the maximum at $\theta  = 0$, i.e. in the direction of the initial
pion momentum. Therefore, it is natural to  place the detector in
this direction from the source in order to register the maximal
possible number of events.

Since the azimuthal angle $\varphi$ is not defined for $\theta  =
0$, in order to find the differential probability $\frac{{dW_1
}}{{\sin \theta d \theta  }}$ at $\theta  = 0$ first  we have to
average the differential probability $\frac{{dW_1 }}{{d\Omega }}$
over the angle $\varphi$ and then to take the limit $\theta
\rightarrow 0$. As a result, we get the following differential
probability of the neutrino production process in the direction of
the initial pion momentum:
\begin{equation}
\left. \frac{{dW_1 }}{{\sin \theta d \theta  }} \right|_{\theta =
0} =  \frac{{G_F^2 \cos ^2 \theta_c\, f_\pi ^2 }}{{8(2\pi)^2
}}\frac{{m_{(\mu )}^2 \left( {m_\pi ^2  - m_{(\mu )}^2 } \right)^2
}}{{p_\pi ^0 \left( {p_\pi ^0  - \left| {\vec p_\pi  } \right|}
\right)^2 }}.
\end{equation}

Let us take a closer look at the registration probability $W_2$. After all the substitutions the
 absolute values and products of the time-dependent factors $A_i$ and $B_i$ defined
 in (\ref{3}) are expressed in the form:
\begin{eqnarray}
 \left| {A_i } \right|^2  &=& \left| {U_{2i} } \right|^2 , \\
 \left| {B_i } \right|^2  &=& \left| {U_{1i} } \right|^2
 \sum\limits_{\scriptstyle k,l = 1 \hfill \atop
  \scriptstyle k < l \hfill}^3 {\left[ {-4{\mathop{\rm Re}\nolimits}
  \left( {U_{1k} U_{1l}^ *  U_{2k}^ *  U_{2l} } \right)\sin^2
  \left( {\frac{{m_k^2  - m_l^2 }}{{4\left| {\vec p} \right|}}L} \right) + } \right.} \nonumber \\
 & & \qquad \qquad \quad \ \left. { + 2 {\mathop{\rm Im}\nolimits}
 \left( {U_{1k} U_{1l}^ *  U_{2k}^ *  U_{2l} } \right)
 \sin \left( {\frac{{m_k^2  - m_l^2 }}{{2\left| {\vec p} \right|}}L} \right)} \right], \\
 {\mathop{\rm Re}\nolimits} \left( {A_i B_i^ *  } \right) &=&
 {\mathop{\rm Re}\nolimits} \left( {U_{1i} U_{2i}^ *
 \sum\limits_{k = 1}^3 {U_{1k}^ *  U_{2k}
  e^{ - i\frac{{m_i^2  - m_k^2 }}{{2\left| {\vec p} \right|}}L} } } \right).
\end{eqnarray}
Substituting these expressions into (\ref{sqr_amp_2}) summed over
$i$ gives:
\begin{eqnarray}
 & & \sum\limits_{i = 1}^3 {\left\langle {\left| {M_2^{(i)} } \right|^2 } \right\rangle }  =
 64G_F^{\, 2} \left\{ {\left[ {2\sin ^2 \theta _W \sum\limits_{\scriptstyle k,l = 1 \hfill \atop
  \scriptstyle k < l \hfill}^3 {\left[ {-4 {\mathop{\rm Re}\nolimits}
  \left( {U_{1k} U_{1l}^ *  U_{2k}^ *  U_{2l} } \right)
  \sin ^2 \left( {\frac{{m_k^2  - m_l^2 }}{{4\left| {\vec p} \right|}}L}
  \right) + } \right.} } \right.} \right. \nonumber \\
 & &  \left. { \left. {\left. { +2{\mathop{\rm Im}\nolimits}
 \left( {U_{1k} U_{1l}^ *  U_{2k}^ *  U_{2l} } \right)
 \sin \left( {\frac{{m_k^2  - m_l^2 }}{{2\left| {\vec p} \right|}}L} \right)} \right] +
 \left( { - \frac{1}{2} + \sin ^2 \theta _W } \right)^2 } \right]\left( {k_1 p} \right)^2  + } \right. \nonumber \\
 & &  \left. { +\sin ^4 \theta _W \left( {kp} \right)^2  - \left[ {\sum\limits_{\scriptstyle k,l = 1 \hfill \atop
  \scriptstyle k < l \hfill}^3 {\left[ {-4 {\mathop{\rm Re}\nolimits}
  \left( {U_{1k} U_{1l}^ *  U_{2k}^ *  U_{2l} } \right)
  \sin ^2 \left( {\frac{{m_k^2  - m_l^2 }}{{4\left| {\vec p} \right|}}L} \right) } \right. + } } \right. } \right.  \nonumber \\
 & &  \left. {\left. {\left. { +2{\mathop{\rm Im}\nolimits}
 \left( {U_{1k} U_{1l}^ *  U_{2k}^ *  U_{2l} } \right)
 \sin \left( {\frac{{m_k^2  - m_l^2 }}{{2\left| {\vec p} \right|}}L} \right)} \right] +
 \left( { - \frac{1}{2} + \sin ^2 \theta _W } \right)} \right]
 \sin ^2 \theta _W m^2 \left( {k_2 p} \right)} \right\}.
\end{eqnarray}
Now one should substitute this expression into (\ref{prob_2})
summed over $i$. Using the formulas for neutrino-electron
scattering kinematics presented in \S 16 of textbook
\cite{Okun:1982ap},  evaluating the integral and substituting
$\left| {\vec p} \right| = \left| {\vec p} \right|^* $ defined in
eq. (\ref{p^star}), we get the following result:
\begin{eqnarray}
 W_2 &=& \frac{{G_F^{\, 2} m}}{{2\pi }}\frac{{2(\left| {\vec p} \right|^*)^2 }}{{2\left| {\vec p} \right|^* + m}}
 \left[ {1 - 2\sin ^2 \theta _W \left( {1 + \frac{{2\left| {\vec p} \right|^*}}{{2\left| {\vec p} \right|^* + m}}} \right) +
 4\sin ^4 \theta _W \left( {1 + \frac{1}{3}\left( {\frac{{2\left| {\vec p} \right|^*}}{{2\left| {\vec p} \right|^* + m}}} \right)^2 } \right) + } \right. \nonumber \\
& &  \left. { +4\sin ^2 \theta _W \left( {1 + \frac{{2\left| {\vec
p} \right|^*}}{{2\left| {\vec p} \right|^* + m}}} \right) \left\{
{ - 4\sum\limits_{\scriptstyle k,l = 1 \hfill \atop
  \scriptstyle k > l \hfill}^3 {\left[ {{\mathop{\rm Re}\nolimits} \left( {U_{1k} U_{1l}^ *  U_{2k}^ *  U_{2l} } \right)\sin ^2
  \left( {\frac{{m_k^2  - m_l^2 }}{{4\left| {\vec p} \right|^*}}L} \right)} \right]}  + } \right. } \right.  \nonumber \\
& & \left. { \left. { + 2 \sum\limits_{\scriptstyle k,l = 1 \hfill
\atop
  \scriptstyle k > l \hfill}^3 {\left[ {{\mathop{\rm Im}\nolimits} \left( {U_{1k} U_{1l}^ *  U_{2k}^ *  U_{2l} } \right)\sin
  \left( {\frac{{m_k^2  - m_l^2 }}{{2\left| {\vec p} \right|^*}}L} \right)} \right] } } \right\} } \right] . \label{W_2}
\end{eqnarray}

In the approximation of massless neutrinos  $\frac{{dW_1
}}{{d\Omega }}$ coincides with the neutrino probability flux and
$W_2$ coincides with the cross section of the scattering process
of a massless fermion on an electron, which can be expressed as
$P_{\mu e}(L) \sigma _{\nu _e e}  + \left( {1 - P_{\mu e}(L) }
\right)\sigma _{\nu _\mu  e}$, where
\begin{eqnarray*}
& & P_{\mu e}(L) = { { - 4\sum\limits_{\scriptstyle k,l = 1
\hfill \atop   \scriptstyle k > l \hfill}^3 {\left[ {{\mathop{\rm
Re}\nolimits} \left( {U_{1k} U_{1l}^ *  U_{2k}^ *
U_{2l} }   \right)\sin ^2 \left( {\frac{{m_k^2  - m_l^2 }}{{4\left| {\vec p} \right|^*}}L} \right)} \right]}  + } } \qquad \qquad \qquad  \\
& & \qquad \qquad \qquad \qquad \ \ + 2 \sum\limits_{\scriptstyle k,l = 1 \hfill \atop
  \scriptstyle k > l \hfill}^3 {\left[ {{\mathop{\rm Im}\nolimits} \left( {U_{1k} U_{1l}^ *  U_{2k}^ *  U_{2l} }
  \right)\sin \left( {\frac{{m_k^2  - m_l^2 }}{{2\left| {\vec p} \right|^*}}L} \right)} \right] }
\end{eqnarray*}
denotes the distance-dependent probability of the transition $\nu
_\mu \to \nu _e$. Thus, we have obtained that the probability of
detecting an electron is equal to the probability of the
production, in the source,  of neutrino with the momentum aimed in
direction of  the detector multiplied by the probability of the
neutrino interaction in the detector, which is expressed in terms
of the muon and electron neutrino interaction cross sections and
the standard distance-dependent $\nu _\mu \to \nu _e$ transition
probability, i.e. we have actually exactly reproduced the result
of the standard approach to neutrino oscillations in the framework
of QFT without making use of the neutrino flavor states and
difficulties associated with applying of wave packets.

Since the incoming $\pi$-mesons always  have a momentum
distribution, the total neutrino probability flux can be obtained
by performing the average of $\frac{{dW }}{{d\Omega }}$ over the
momentum distribution of the incoming $\pi$-mesons. In this case
the magnitude of the momentum of the virtual neutrinos is not
fixed, which results in the blurring of the interference pattern
and gives rise to the corresponding coherence length. The number
of events in the detector per unit time can be found by
integrating the corresponding probability and the densities of
$\pi$-mesons and electrons over the volumes of the neutrino source
and detector.

\section{Oscillations in experiments with neutrino detection by charged-current interactions only}
Let us consider the process, where a neutrino is produced in the
muonic decay of  $\pi^+$-meson, as in the previous case, but it is
detected only by the charged-current interaction with a nucleus.
The process is described in the lowest order by the diagram:
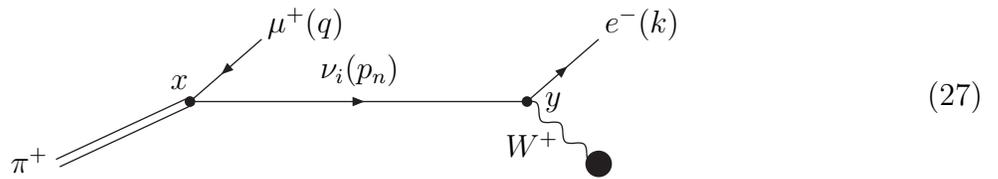
\vspace{1cm}
\begin{center}
\begin{picture}(193,81)(0,0)
\Text(70.0,94.0)[l]{$\mu^+ (q)$}\ArrowLine(67.5,88.0)(40.5,64.5)
\Text(40,72)[r]{$x$} \Line(-10,42.7)(40.5,66.2) \Line(-8.6,39.9)(42,63.5)
\Text(-13,42.5)[r]{$\pi^+$}
\Vertex(40.5,64.5){2} \ArrowLine(40.5,64.5)(167.5,64.5) \Vertex
(167.5,64.5){2} \Text(104.8,70.5)[b]{$\nu_i ( p_n )$}
\ArrowLine(167.5,64.5)(194.5,88.0) \Text(197.5,94.0)[l]{$e^- ( k )$}
\Text(175.0,64.5)[l]{$y$} \Text(160.0,48.5)[l]{$W^+$}
\Photon(167.5,64.5)(194.5,41.0){2}{3.0} \Vertex (194.5,41.0){5}
\Text(330.0,60.5)[b]{\addtocounter{equation}{1}(\arabic{equation})}
\setcounter{diags1}{\value{equation}}
\end{picture}
\end{center}
\vspace{-1cm}
which should be summed over the type $i=1,2,3$ of the intermediate
neutrino mass eigenstate. The filled circle stands for the matrix
element $j_\mu$ of the charged weak hadron current. Since the
neutrino energy in the muonic decay of pion is of the order of 30
MeV, the interaction of the virtual neutrinos with a nucleus can
result in the disintegration of the latter. To be specific, we
will consider first only the two body final state and suppose that
an initial nucleus $^{A}_{Z} X$ absorbs $W^+$-boson and turns into
the final nucleus $^{A}_{Z + 1} X$, thus
$$j_\mu = \left <^{A}_{Z + 1} X \right| j_\mu^{(h)} \left| ^{A}_{Z} X \right>.$$

Using again the approximation of Fermi's interaction one  can
write out the amplitude in the momentum representation
corresponding to diagram (\arabic{diags1}) summed over all three
neutrino mass eigenstates in the case, where the time difference
$y^0-x^0$ between the production and detection points is fixed and
equal to $T$:
\begin{equation}\label{amp_cc}
M =  - i\frac{{G_F^{\, 2} }}{{2p_n^0 }}\cos \theta_c\, f_\pi
\varphi _\pi  m_{(\mu )} \sum\limits_{i = 1}^3 {U_{1i} U_{2i}^ *
e^{ - i\frac{{m_i^2  - p^2 }}{{2p_n^0 }}T} } j_\mu \bar u\left( k
\right)\gamma ^\mu  \hat p_n \left( {1 + \gamma ^5 }
\right)\upsilon \left( q \right).
\end{equation}
Here the particle 4-momenta are defined similarly to the previous
section, as it is shown in the diagram.

The squared amplitude averaged with respect to the incoming
particles polarizations and summed over the outgoing particles
polarizations factorizes as follows:
\begin{eqnarray}
 \left\langle {\left| M \right|^2 } \right\rangle  &=&
 \left\langle {\left| {M_1 } \right|^2 } \right\rangle
 \left\langle {\left| {M_2 } \right|^2 } \right\rangle
 \frac{1}{{4\left( {p_n^0 } \right)^2 }} \times \label{sqr_amp_cc_tot} \\
  & & \! \! \! \! \! \! \! \! \! \! \! \! \! \! \! \! \! \! \! \! \! \! \! \! \!
    \times \sum\limits_{\scriptstyle i,k = 1 \hfill \atop
  \scriptstyle i < k \hfill}^3 {\left[ {-4{\mathop{\rm Re}\nolimits}
  \left( {U_{1i} U_{1k}^ *  U_{2i}^ *  U_{2k} } \right)
  \sin^2 \left( {\frac{{m_i^2  - m_k^2 }}{{4p_n^0 }}T} \right) +
  2 {\mathop{\rm Im}\nolimits} \left( {U_{1i} U_{1k}^ *  U_{2i}^ *  U_{2k} } \right)
  \sin \left( {\frac{{m_i^2  - m_k^2 }}{{2p_n^0 }}T} \right)} \right]} , \nonumber
 \end{eqnarray}
where $\left\langle {\left| {M_1 } \right|^2 } \right\rangle$  is
the squared amplitude of the pion decay into  antimuon and a
massless fermion, given by formula (\ref{sqr_amp_1}), and
\begin{equation} \label{sqr_amp_cc_2}
 \left\langle {\left| {M_2 } \right|^2 } \right\rangle  =
 4G_F^{\, 2} \left[ {k^\mu  p_n^\nu   + k^\nu  p_n^\mu   -
 \left( {p_n k} \right)g^{\mu \nu }  +
 i\varepsilon ^{ \mu \nu \alpha \beta } k_\alpha  p_{n\beta } } \right] \left(W_{\mu \nu }^{(S)} + i W_{\mu \nu
 }^{(A)}\right)
\end{equation}
is the squared amplitude of the scattering process of the initial
nucleus and a massless fermion resulting in the production of the
final nucleus and an electron. Here the nuclear tensor $W_{\mu \nu
} = W_{\mu \nu }^{(S)} + i W_{\mu \nu }^{(A)} = \left\langle
{j_\mu j_\nu ^ +  } \right\rangle$ characterizes the interaction
of the nucleus with a virtual $W^+$-boson, its symmetric part
$W_{\mu \nu }^{(S)}$ being real and antisymmetric part $i W_{\mu
\nu }^{(A)}$ being imaginary.

Let us denote the 4-momentum of  $\pi^+$-meson again  by $p_\pi$
and the 4-momenta of the initial and final nuclei by $P = (E, \vec
P), \ P^{\prime} = (E^{\prime}, \vec P^{\prime})$, respectively.
Following the  outlined recipe, in order to find the probability
of the process one has to multiply the amplitude $\left\langle
{\left| M \right|^2 } \right\rangle$ by the delta function of the
energy-momentum conservation $(2\pi)^4 \delta ( p_\pi + P - q - k
- P^{\prime})$ and by the delta function $2\pi \delta ( p_\pi - q
- p)$, which fixes the momentum of the intermediate neutrinos, to
substitute $p$ instead of $p_n$ and to integrate with respect to
the momentum of the final particles. We  may also replace the time
interval $T$ by ${{Lp^0 } \mathord{\left/
 {\vphantom {{Lp^0 } {\left| {\vec p} \right|}}} \right.
 \kern-\nulldelimiterspace} {\left| {\vec p} \right|}}$ to pass
 from the time-dependent factor to the distance-dependent factor,
 because the momentum $p_n$ is now selected to be equal to $p$.
 The result has to be integrated with respect to $\left| {\vec p} \right|$,
 and this integration is performed using the additional delta function.
 As a result of all these transformations we have:
\begin{eqnarray}
& & \! \! \! \! \! \! \! \! \! \!  \frac{{dW}}{{d \Omega}} = \int { \frac{{dW}}{{d\vec p}}  \left| {\vec p} \right|^2 d\left| {\vec p} \right| } = \frac{{dW_1}}{{d \Omega}} \left. {W_2 } \right|_{\left| {\vec p} \right| = \left| {\vec p} \right|^ *  }  \times \label{prob_tot_cc} \\
 & & \! \!  \times \sum\limits_{\scriptstyle i,k = 1 \hfill \atop
  \scriptstyle i < k \hfill}^3 {\left[ {-4{\mathop{\rm Re}\nolimits}
  \left( {U_{1i} U_{1k}^ *  U_{2i}^ *  U_{2k} } \right)
  \sin^2 \left( {\frac{{m_i^2  - m_k^2 }}{{4\left| {\vec p} \right|^*}}L} \right) +
  2 {\mathop{\rm Im}\nolimits} \left( {U_{1i} U_{1k}^ *  U_{2i}^ *  U_{2k} } \right)
  \sin \left( {\frac{{m_i^2  - m_k^2 }}{{2\left| {\vec p} \right|^*}}L} \right)} \right]} , \nonumber
\end{eqnarray}
where ${\left| {\vec p} \right|^ *  }$ is given by (\ref{p^star}),
$\frac{{dW_1}}{{d \Omega}}$ stands for the differential
probability of the $\pi$-meson decay into an antimuon and a
massless fermion with the fixed direction of the momentum, given
by (\ref{dif_prob_prod}), and
\begin{eqnarray}
W_2  = \frac{1}{{2p^0 2E}}\int {\frac{{d^3 k}}{{\left( {2\pi } \right)^3 2k^0 }}}
\frac{{d^3 P'}}{{\left( {2\pi } \right)^3 2E'}}\left\langle {\left| {M_2 } \right|^2 } \right\rangle
\left( {2\pi } \right)^4 \delta \left( {P + p - P' - k} \right)
\end{eqnarray}
is the probability of the scattering process of a nucleus and a
massless fermion resulting in the production of the final nucleus
and electron. In fact, this probability should be replaced by the
probability of the inclusive scattering process, where only the
final electron is detected. However, this does not influence the
result that the factor in formula (\ref{prob_tot_cc}) exactly
coincides with the one we expect  for the $\nu_\mu \to \nu_e$
transition probability in the conventional approach. The
number of events in the detector can be found exactly in the same
way, as it was explained in the end of the previous section.

\section{Conclusion}
In the present paper we have shown that the lepton flavor changing
neutrino oscillation processes can be consistently described in
quantum field theory using only plane wave states of the
involved particles. In the framework of the Standard Model
minimally extended by the right neutrino singlets we have used the
modified perturbative formalism put forward in paper
\cite{Volobuev:2017izt} and developed in paper
\cite{Egorov:2017qgk}. It is based on the conventional S-matrix
approach supplemented by the modified rules of passing from the
coordinate representation to the momentum representation. These
rules allow us to construct the modified Feynman propagator in the
momentum representation corresponding to the experimental
situation at hand, which we call the time-dependent propagator.
Unlike the standard S-matrix formalism, our approach is adequate
for describing the processes passing at finite distances and
finite time intervals. The calculations
 are simple and very similar to those in the standard
perturbative S-matrix formalism in the momentum representation.
The modified S-matrix formalism is physically transparent and has
the advantage of not violating energy-momentum conservation. It is
important to note that we do not make use of the neutrino flavor
states in the model, working only with the neutrino mass
eigenstates.

This technique has been used for calculating the oscillating
probabilities of the processes, where neutrinos are produced in
the muonic decay of $\pi^+$-meson and detected in the
neutral-current and charged-current interactions with electrons or
just the charged-current interaction with nuclei. It was
explicitly shown that the approach exactly reproduces the results
of the standard formalism.

\bigskip
{\large \bf Acknowledgments}
\medskip \\
\noindent The authors are  grateful to E. Boos,  A. Lobanov and M.
Smolyakov for reading the manuscript and making important comments
and to L. Slad for useful discussions.   Analytical calculations
of the amplitudes have been carried out with the help of  the
COMPHEP and REDUCE packages. The work of V. Egorov was supported
by the Foundation for the Advancement of Theoretical Physics and
Mathematics ``BASIS''.

\end{document}